\documentclass[a4paper,twocolumn,11pt, accepted=2026-05-04]{quantumarticle}
\pdfoutput=1
\usepackage[utf8]{inputenc}
\usepackage[english]{babel}
\usepackage[T1]{fontenc}
\usepackage{amsmath}
\usepackage{amssymb}
\usepackage{hyperref}
\usepackage{graphicx}
\usepackage{float}
\usepackage{subfigure}  
\usepackage{amsthm}
\usepackage{mathrsfs}
\usepackage{amsfonts}
\usepackage{physics}
\usepackage{braket}
\usepackage{tikz}
\usepackage{todonotes}
\usepackage{lipsum}
\usepackage{parskip}
\newcommand*{\centerfloat}{%
  \parindent \z@
  \leftskip \z@ \@plus 1fil \@minus \textwidth
  \rightskip\leftskip
  \parfillskip \z@skip}

\usepackage[deletedmarkup=sout, addedmarkup=bf]{changes}

\definechangesauthor[name={Author}, color=red]{del}
\setdeletedmarkup{\textcolor{red}{\sout{#1}}}

\begin{document}

\title{Catalytic entanglement transformations with noisy hardware}

\author[1,2]{Hemant Sharma}
\email{h.sharma-1@tudelft.nl}
\orcid{0000-0002-4035-4300}
\author[4]{Aleksandr Mokeev}
\author[2]{Jonas Helsen}
\author[3]{Johannes Borregaard}

\affil[1]{QuTech, Delft University of Technology, 2628 CJ, Delft, The Netherlands} 
\affil[2]{QuSoft and CWI, Science Park 123, 1098 XG Amsterdam, The Netherlands}
\affil[3]{Department of Physics, Harvard University, Cambridge, Massachusetts 02138, USA}
\affil[4]{Department of Electrical Engineering, Eindhoven University of Technology, 5600 MB, Eindhoven, The Netherlands}

\newcommand{\hs}[1]{{\color{orange} \;HS:#1}}
\newcommand{\jb}[1]{{\color{blue} \;JB:#1}}


\maketitle
\begin{abstract}
The availability of certain entangled resource states (catalyst states) can enhance the rate of converting several less entangled states into fewer highly entangled states in a process known as catalytic entanglement concentration (EC). Here, we extend catalytic EC from pure states to mixed states and numerically benchmark it against non-catalytic EC and distillation in the presence of state-preparation errors and operational errors. Furthermore, we analyse the re-usability of catalysts in the presence of such errors. To do this, we introduce a novel recipe for determining the positive-operator valued measurements (POVM) required for EC transformations, which allows for making tradeoffs between the number of communication rounds and the number of auxiliary qubits required. We find that in the presence of low operational errors and depolarising noise, catalytic EC can provide better rates than distillation and non-catalytic EC.

\end{abstract}
\section{Introduction}
Entanglement is a crucial resource for building quantum networks for secure communication~\cite{Pirandola_2020, quantum_internet}, distributed quantum computing~\cite{dstcomp}, and networked quantum sensing~\cite{Khabiboulline_2019}. However, the distribution of entanglement in a network is challenging because of errors and loss, which deteriorate the quality of shared entanglement. Several protocols have been developed to convert low-quality entanglement into high-quality entanglement, including distillation~\cite{Bennett_1996, Deutsch_1996} and entanglement concentration (EC)~\cite {ent_cont, lo_1999}.

Bilocal Clifford distillation protocols (or simply distillation) have been the canonical approach for creating high-quality entanglement from mixed entangled states due to their relatively simple execution. They have been studied intensely~\cite{Rozp_dek_2018, Krastanov_2019}, and have been demonstrated in experiments~\cite{meesala2023, pompili, experimental_implementation}. EC, on the other hand, has not been demonstrated experimentally. In particular, previous theoretical studies have considered only pure states, which makes them incompatible with realistic, noisy hardware. Moreover, they have been studied only in the absence of operational errors~\cite{ent_conc_bo, Duan_2001}. This includes entanglement catalysis protocols that we treat as an extension of EC. Entanglement catalysis uses pre-shared entanglement that remains unchanged after the protocol to increase the rate of EC~\cite{santra, Datta_2024}. It can also be used for catalytic quantum teleportation~\cite{new_cite, cat_teleportation}.

In this paper, we extend the analysis of EC protocols to the more practically relevant regime of noisy hardware. We do this by considering both mixed state inputs and operational errors.
\begin{figure*}[t]
 \begin{minipage}{\textwidth}
 \includegraphics[width=\linewidth]{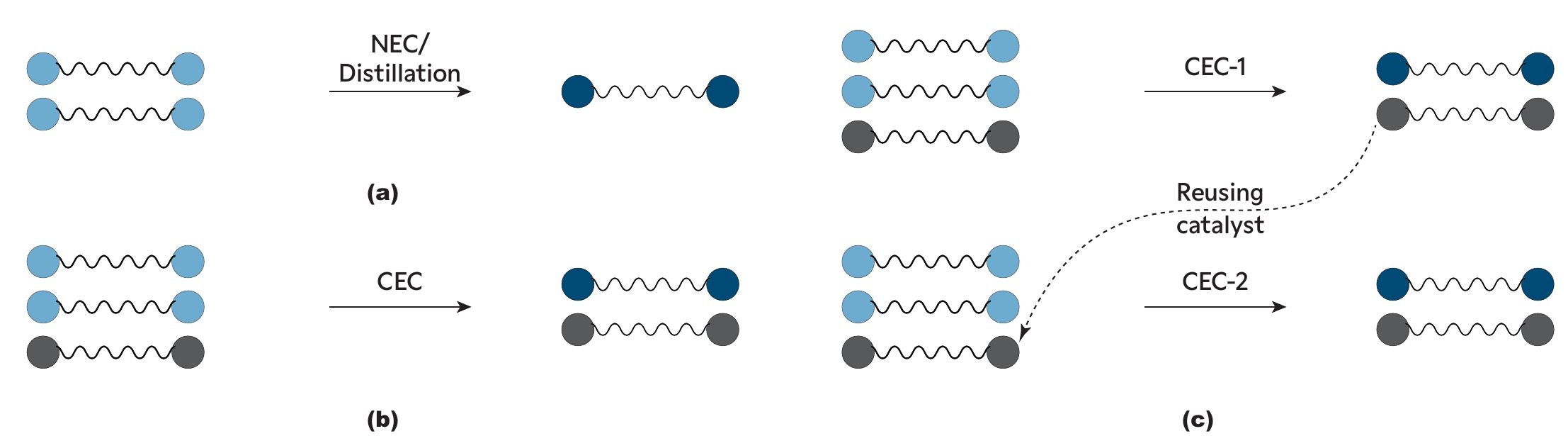}
 \end{minipage}
 \caption{In distillation and non-catalytic EC, we consume two less entangled states to create a more entangled state, as shown in (a). In case of catalytic EC (b), we also consume two less entangled states but make use of a third catalyst state in the process. Thus, three initial states are converted into one highly entangled state and the catalyst state. To study the reuse of a catalyst state, we consider that a new catalyst is first used for catalytic EC. It is then reused for catalytic EC of another set of initial states with the same amount of state preparation error.}
 \label{fig: setup}
\end{figure*}
As part of our analysis, we introduce a novel recipe for determining the local operations and classical communication needed for performing the EC protocols. In particular, we allow for a practical trade-off between the number of communication rounds and the number of auxiliary qubits in the implementation based on hardware constraints.

We numerically simulate EC both with and without entanglement catalysis and compare the performances to a canonical protocol of distillation in the presence of both state-preparation errors and operational errors. Our current numerical analysis considers bipartite states but our general method can be extended to multipartite states following the extension in Ref.~\cite{Xin_2007}. We find that the distillation protocol is more resistant against incoherent, depolarising-type errors, whereas the EC protocols are more resilient against coherent, unitary-type errors. For operational errors, we find that they affect both catalytic EC and non-catalytic EC more severely than distillation due to the relatively complex circuits needed for performing the required measurements. However, for operational errors $\lesssim0.1$\%, we find that catalytic EC results in both higher fidelity output states and succeeds with a higher success probability than non-catalytic protocols and distillation. Consequently, entanglement catalysis seems suited for the concentration of entanglement between logical qubits utilising quantum error correction to suppress operational errors as recently envisioned for fault-tolerant modular quantum computation~\cite{pattison2024,ataides2025}.

\section{Execution of entanglement concentration}\label{sec: locc_protocol}
We consider the conversion of two less entangled states (referred to as the prepared states) into one more entangled output state using EC protocols and distillation (see Fig.~\ref{fig: setup}). For catalytic EC, we assume that we have access to a previously prepared error-free optimal catalyst state to assist the process. The catalyst state should be returned at the end of the protocol (see Fig.~\ref{fig: setup}(b)). A catalyst state is considered optimal if it maximises the conversion probability over all possible catalyst states~\cite{santra}. In case of distillation, we convert the two copies of prepared states into a more entangled output state using optimal 2-1 distillation protocols as described in Ref.~\cite{Krastanov_2019}.

\begin{figure*}[t]
 \centering
 \begin{minipage}{0.99\textwidth}
 \includegraphics[width = 0.95\linewidth]{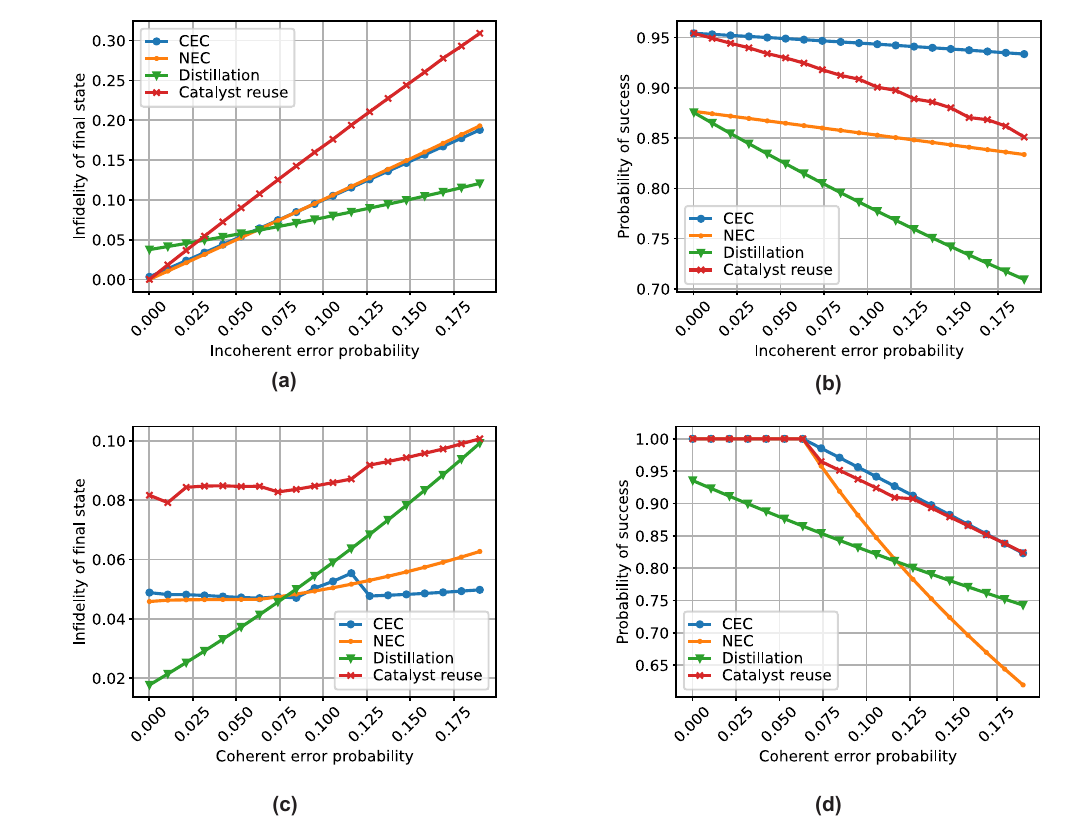}
 \end{minipage}
 \caption{Here, we plot the performance of catalytic EC (CEC), non-catalytic EC (NEC), distillation and when reusing the catalyst (called "catalyst reuse"). For increasing incoherent error probability and constant coherent error probability ($a=0.9$), we plot the infidelity and success probability in (a) and (b), respectively. In (c) and (d), we plot infidelity and success probability, respectively, for increasing coherent error probability and constant incoherent error probability ($p=0.95$). From (a) and (b), we deduce that the infidelity of the final state obtained from catalytic EC and non-catalytic EC is dependent on the amount of incoherent error, while the success probability remains relatively constant. From (c) and (d), we note that the success probability is dependent on the coherent error while the infidelity is nearly unaffected. We have assumed equal weight error terms for both incoherent and coherent errors ($\epsilon_k=\varepsilon_k^2=1/3$ for $k\in\{x,y,z\}$).}
 \label{fig: depol_error}
\end{figure*}

When reusing the catalyst state, we consider an optimal catalyst that might be deteriorated from being used once for the conversion of a pair of prepared states. We then use the deteriorated catalyst state to assist a second round of catalytic EC  using another set of prepared states, as shown in Fig.~\ref{fig: setup}(c).

For EC, we implement the protocol proposed by Vidal for entanglement transformations in Ref.~\cite{Vidal_1999}. This protocol has two parts: (1) an LOCC conversion from the given initial state to an intermediate state, followed by (2) an SLOCC conversion from the intermediate state to the final state. Here, LOCC refers to Local Operations and Classical Communication, while SLOCC refers to Stochastic Local Operations and Classical Communication. The former is the class of operations that can be implemented by deterministic local operations on each of the two systems of the entangled states together with classical communication between the systems, while the latter includes non-deterministic (stochastic) local operations (for further details, we refer to Appendix~\ref{sec: locc_intro}). Ref.~\cite{Vidal_1999} provided the recipe for determining the operations required for the SLOCC part of the protocol in a gate-based quantum processor. For the LOCC part, we introduce a new way to find the required operations by combining two previously proposed methods~\cite{Nielsen_1999,simple_locc} for LOCC conversions, allowing for a trade-off between the required number of auxiliary qubits at each system and rounds of classical communication between them. We discuss this further in Section~\ref{sec: oper_err} when analysing the effect of operational errors.

The protocol proposed in Ref.~\cite{Vidal_1999} is a \emph{conclusive} protocol~\cite{Vidal_2000}. A conclusive protocol converts a given initial state into a desired final state with unit fidelity and finite probability using SLOCC. To adapt EC to mixed states, we consider the pure state that corresponds to the largest eigenvalue of the mixed initial state. We then use this pure state as the surrogate to find the necessary operations for EC and apply these operations to the prepared mixed initial state.

\begin{figure*}[t]
 \centering
 \begin{minipage}{0.99\textwidth}
 \includegraphics[width = 0.95\linewidth]{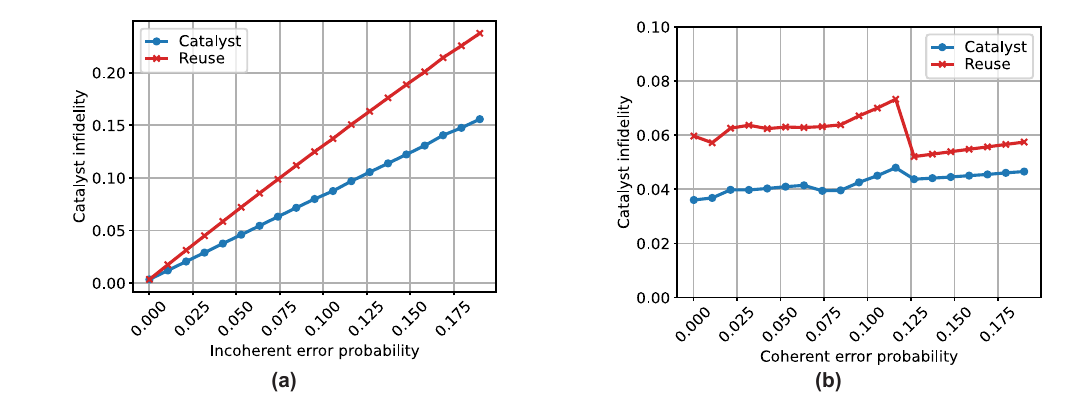}
 \end{minipage}
 \caption{We plot the infidelity of the catalyst state before and after it is used for catalytic EC, quantifying the change in the state of the catalyst. The fidelity of the catalyst state is highly affected by the amount of incoherent error(a) while the effect of coherent errors is limited. We have assumed equal weight error terms for both incoherent and coherent errors ($\epsilon_k=\varepsilon_k^2=1/3$ for $k\in\{x,y,z\}$).}
 \label{fig: cat_perf}
\end{figure*}

\section{Results and discussion}
During distillation and EC, different forms of errors can affect the performance of the protocol. We divide the errors into state preparation errors and operational errors. Operational errors affect the protocols during the application of single and multi-qubit gates. We further divide state preparation errors into incoherent (depolarising) errors and coherent (unitary) errors. 

\subsection{State preparation errors}\label{sec: state_prep}
Recent experimental demonstrations of entanglement distribution with Silicon vacancy centres (SiV)~\cite{Bersin_2024, Bhaskar_2020, knaut2024}, Nitrogen vacancy centres~\cite{bradley2021robustquantumnetworkmemorybased, Stolk_2024}, and neutral atoms~\cite{Thomas_2022, Thomas_2024} have various imperfections that limit the achievable entanglement fidelity. For example, demonstrated entanglement distribution with SiVs suffers from both incoherent errors from qubit decoherence as well as coherent errors from imperfect operations and limited contrast of the spin-dependent optical reflection mediating the spin-photon entanglement generation~\cite{Bhaskar_2020, knaut2024}. To encompass both types of error, we introduce two error channels to model general state preparation errors. The first channel parametrises the effect of incoherent errors modelled as a general depolarising channel acting on the state with error probability, $p_{d}$:
\begin{eqnarray}\label{eq: depol}
\Delta_p(\rho) &=& (1-p_{d}) I\rho I \nonumber \\
&&+ p_{d}\big(\epsilon_x X \rho X +\epsilon_z Z \rho Z + \epsilon_y Y\rho Y \big),
\end{eqnarray}
where $\{\epsilon_{x},\epsilon_{z},\epsilon_{y}\}$ are the relative weight of the error terms ($\sum_{k=\{x,y,z\}}\epsilon_{k}=1$). For a completely depolarizing channel $\epsilon_{x}=\epsilon_{z}=\epsilon_{y}=1/3$.

The second channel parametrises the effect of coherent errors on the state and is expressed as:
\begin{eqnarray}
\ket{{\Phi}_a} &=& \sqrt{(1-a)} \ket{\Phi^+} \nonumber \\
&&+\sqrt{a} \left [ \varepsilon_z \ket{\Phi^-}+ \varepsilon_x\ket{\Psi^+}+ \varepsilon_{y}\ket{\Psi^-} \right ],\quad
\end{eqnarray}
where $a$ is the error probability and $\{\varepsilon_{x},\varepsilon_{z},\varepsilon_{y}\}$ are the relative amplitudes of the error terms ($\sum_{k=\{x,y,z\}}|\varepsilon_{k}|^2=1$). We use the notation $\{\ket{\Phi^{\pm}},\ket{\Psi^{\pm}}\}$ to denote the states of the standard two-qubit Bell basis where $\ket{\Phi^{\pm}}=(\ket{0,0}\pm\ket{1,1})/\sqrt{2}$ and $\ket{\Psi^{\pm}}=(\ket{0,1}\pm\ket{1,0})/\sqrt{2}$. Without loss of generality, we will always assume that the desired entangled state is $\ket{\Phi^+}$. We consider that the prepared states have both coherent and incoherent errors resulting in states of the form $\rho(a,p_{d}) = \Delta_{p_{d}}(\ketbra{\Phi_a})$. For extending EC to mixed states of this form, we find a surrogate state that is used to find the catalyst state (for catalytic EC) and the required operations. This surrogate state is the maximum weight eigenstate of $\rho(a,p_{d})$.

\begin{figure*}[t]
 \centering
 \begin{minipage}{0.99\textwidth}
 \includegraphics[width=0.95\linewidth]{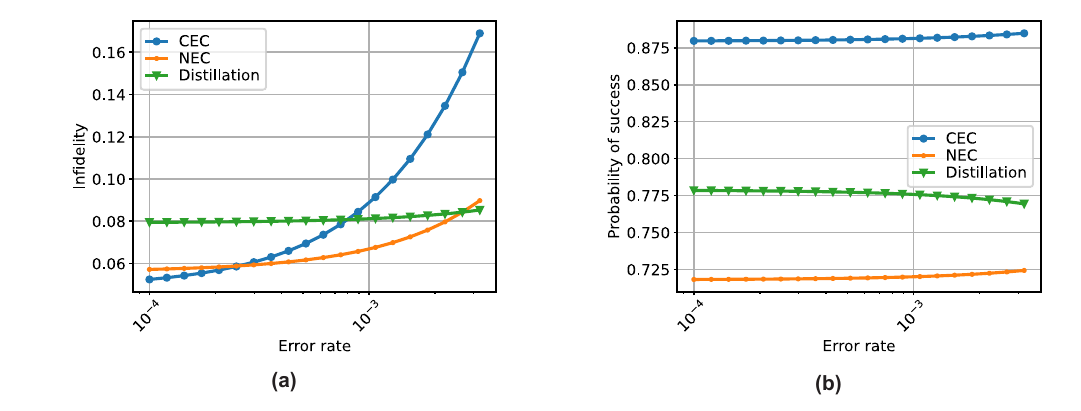}
 \end{minipage}
 \caption{For a given initial state, we plot the infidelity and success probability for increasing operational error for catalytic EC (CEC), non-catalytic EC (NEC) and distillation. The initial state is prepared with $a=0.85$ and $p=0.95$ and equal weight error terms ($\epsilon_k=\varepsilon_k^2=1/3$ for $k\in\{x,y,z\}$). Fig. (a) shows that the infidelity Catalytic EC is most affected by operational errors. It is followed by non-catalytic EC and distillation. This is due to the high number of MCX gates required per round of communication. In (b), we plot the probability of success of our protocols. Distillation's success probability decreases for higher values of operational error, while for others it slightly increases. Therefore, in the presence of higher operational errors, catalytic EC and non-catalytic EC output worse states with higher probability.}
 \label{fig: oper_er}
\end{figure*}

We plot the infidelity of the output state and the probability of success for increasing depolarising and coherent errors in Fig.~\ref{fig: depol_error}, assuming error terms of equal weight for both the incoherent and coherent error channels. Moreover, we assume that the gates required during the conversion are perfect. From Fig.~\ref{fig: depol_error}(a), we find that the infidelity of catalytic EC and non-catalytic EC protocols increases linearly with the depolarisation probability as $f\approx 1-p_d$ for both catalytic and non-catalytic EC. Interestingly, this is different from the naive lower bound of $f \geq 1-2p_{d}$ that one obtains by only considering the highest order term of $\rho(a,p_{d})\otimes\rho(a,p_{d})\approx(1-2p_d)\ket{\Phi_a}\bra{\Phi_a}\otimes \ket{\Phi_a}\bra{\Phi_a} + O(p_d)$ in the limit of $p_d\ll1$. Thus, while the protocol of Ref.~\cite{Vidal_1999} is a conclusive protocol, meaning that the surrogate state $\ket{\Phi_a}$ is converted to the target Bell state ($\ket{\Phi^+}$) with unit fidelity (in the absence of operational errors), it also manages to correct a significant fraction of the higher-order error terms.

From Fig.~\ref{fig: depol_error}(c), we see that coherent errors have little effect on the fidelity of the output states from non-catalytic EC and catalytic EC. The output fidelity of catalytic EC or non-catalytic EC is therefore mainly determined by the probability of incoherent errors. However, from Fig.~\ref{fig: depol_error}(d), we see that the probability of success depends strongly on the amount of coherent error for catalytic EC and non-catalytic EC. The success probability remains at unity for small coherent errors, consistent with results from Ref.~\cite{santra}. This is because for a small amount of coherent error, the surrogate state can be converted into a Bell state deterministically using LOCC. For larger amounts of coherent error, this is not possible, and SLOCC is required, which results in a finite success probability. When comparing to distillation, we see that catalytic EC or non-catalytic EC can lead to both higher output fidelity and larger success probabilities than distillation for small incoherent error probabilities and for large coherent error probabilities.

\subsection{Reusing the catalyst state}

Reusing the catalyst results in a higher probability of success compared to only non-catalytic EC, but it has a higher infidelity compared to both non-catalytic EC and catalytic EC. In Fig.~\ref{fig: cat_perf}, we plot the fidelity of the catalyst state before and after it is used for catalytic EC. We find that increasing the depolarisation error decreases the fidelity of the catalyst state, while increasing the coherent error does not significantly affect the fidelity. Moreover, reusing the catalyst state multiple times results in a higher reduction in the catalyst fidelity due to increased depolarisation.

\subsection{Operational errors}\label{sec: oper_err}

Imperfect gates can lead to errors during the implementation of distillation or EC protocols. Here, we study the effects of these errors on the probability of success and the fidelity of the resulting state.

For the case of EC protocols, we use Naimark's dilation to convert the required POVMs into projective measurements by introducing auxiliary qubits. This allows the execution of these POVMs on a circuit-based quantum system. This is done by first embedding the POVM set into a unitary, called the \textit{embedding unitary}. This unitary acts on both the qubits corresponding to the prepared state and auxiliary qubits. The auxiliary qubits are then measured in the computational basis, implementing a POVM on the prepared state qubits. The number of auxiliary qubits required for implementing the POVMs grows logarithmically with the number of elements in the POVM.

Previous work on constructing these POVMs includes decomposing a doubly-stochastic matrix obtained from the initial and final states. The method proposed by Nielsen~\cite{Nielsen_1999} decomposes a doubly-stochastic matrix into a set of T-transforms and converts them into POVMs with only two elements. However, it requires a number of communication rounds equal to the number of T-transforms. By contrast, the method introduced by Jensen and Schack~\cite{simple_locc}, decomposes the doubly-stochastic matrix into a convex combination of permutation matrices. These permutation matrices are then used to obtain a POVM, which can be implemented in a single round of communication at the cost of additional auxiliary qubits.

In this work, we propose a construction that interpolates between these two approaches. By grouping multiple T-transforms to obtain a single effective POVM, one can reduce the number of communication rounds at the expense of increasing the number of POVM outcomes and hence the number of auxiliary qubits required. Varying the degree of grouping allows one to continuously tune between the two limiting cases represented by Refs.~\cite{Nielsen_1999} and~\cite{simple_locc}.

This flexibility makes it possible to adapt the conversion protocol to different physical constraints. For example, when communication latency is large compared to qubit coherence times, it may be advantageous to reduce the number of communication rounds by using more auxiliary qubits. Conversely, when auxiliary qubits are scarce but communication is fast and reliable, operating closer to the sequential T-transform regime may be preferable. The details of this interpolated construction and the resulting resource trade-offs are discussed in Appendices~\ref{sec: our_method} and~\ref{sec: povm_impl}.



To study the effect of gate errors, we compile the embedding unitaries involved during the EC protocols. Each communication round during a protocol requires the application of such a unitary and measurement of the auxiliary qubits. The structure of this unitary allows us to synthesise it into several multi-controlled unitaries. These multi-controlled unitaries can be implemented using multiple single-qubit gates and at most two multi-controlled X (MCX) gates by extending the circuits shown for CX gates in Ref.~\cite{Barenco_1995, Shende_2006} to MCX gates. Synthesis into MCX gates allows for potentially a more efficient (fewer gates) synthesis compared to if only CX gates were used instead. Moreover, they are native to some quantum computing platforms like neutral atoms~\cite{cczgate, Dlaska_2022, Bluvstein_2023} and trapped ions~\cite{shapira2023fastdesignscalingmultiqubit, shapiratrappedion}. In Appendix~\ref{sec: povm_synthesis}, we discuss the synthesis of the embedding unitary into MCX gates.

For modelling operational errors, we consider that each gate depolarises the qubits it acts on upon execution. The depolarisation error is parametrised by the depolarisation probability ($p_\text{d}$) with equal weight error terms, given in Eq.~(\ref{eq: depol}). In practice, the infidelity of single-qubit gates is often an order of magnitude lower than the error introduced by multi-qubit gates~\cite{cczgate}. Therefore, we ignore the contribution of single-qubit gate errors towards operational errors. Moreover, we assume that during POVM measurement and distillation protocol all projective measurements are error-free and the only sources of errors are gate errors. 

\begin{figure}[t]
 \centering
 \begin{minipage}{0.4\textwidth}
 \includegraphics[width=0.85\linewidth]{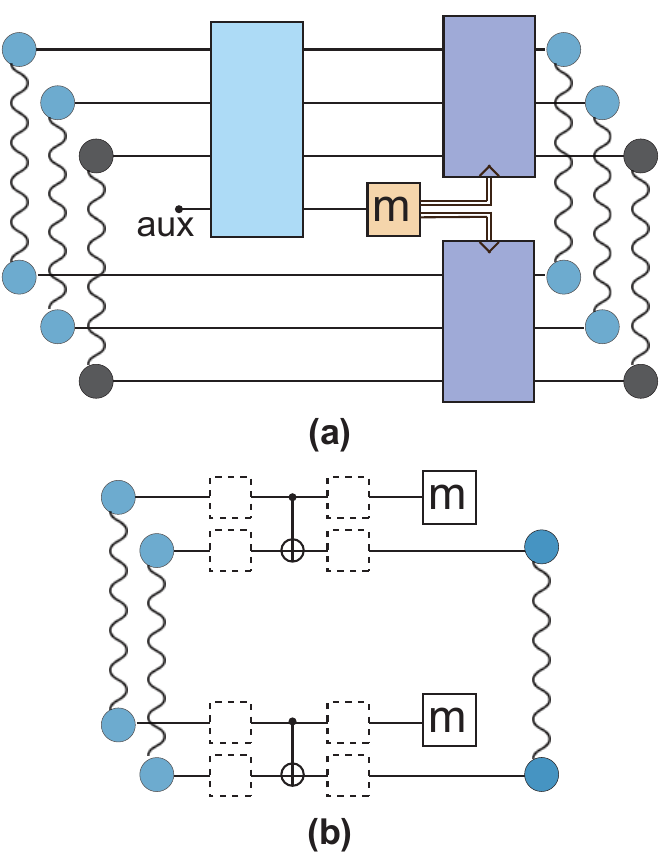}
 \end{minipage}
 \caption{Comparison of circuits for catalytic EC and distillation. (a) shows the circuit for a single round of LOCC during catalytic EC. We note that catalytic EC involves multiple such rounds, with each round involving multi-qubit unitaries before and after measurements. In (b), we show the circuit for distillation. Distillation involves the application of single-qubit Clifford gates and the CNOT gate, followed by measurement in X, Y, or Z bases.}
 \label{fig: distillation}
\end{figure}

The effects of operational error on the performance of catalytic EC, non-catalytic EC and distillation are shown in Fig.~\ref{fig: oper_er}. We note that distillation is relatively robust against operational errors since its circuit only contains a single multi-qubit gate at each system, as shown in Fig.~\ref{fig: distillation}. On the other hand, the fidelity of the state obtained from catalytic EC decreases rapidly compared to both non-catalytic EC and distillation. This is because of the high number of multi-qubit gates required to implement the POVM-embedded unitary on the data qubits and auxiliary qubits. However, the probability of success of both catalytic EC and non-catalytic EC increases slightly as the rate of depolarisation increases. We attribute the increase in the probability of success to the increase in the depolarisation of the output states caused by increasing operational error. Thus, at a higher operational error rate, EC protocols output worse states (lower fidelity) with a higher probability. This suggests that operational errors are detrimental to the performance and reliability of EC protocols. Distillation, on the other hand, outputs states with only slightly worse fidelity and with a lower probability of success, demonstrating its robustness.
Our current analysis is limited to only two elements in the POVM set for implementing LOCC, thus requiring a larger number of communication rounds. However, one can easily find POVM sets such that fewer communication rounds are required by increasing the number of elements in each POVM set, as discussed in Appendix~\ref {sec: our_method}. For implementing POVM sets with a higher number of elements, the embedding unitary would be composed of multi-controlled multi-target (MCMT) unitaries. Such an MCMT unitary can be synthesised into MCX gates and single qubit unitaries using the ideas proposed in Ref.~\cite{Shende_2006, tucci1998}. Through these ideas, we expect the number of MCX gates per round to grow linearly with the number of elements in the POVMs for that round. Therefore, using our method for finding POVMs, one could decide on the size of the POVM set and number of communication rounds depending on the intensity of operational errors and qubit depolarisation during classical communication, such that the overall error is minimised.


\section{Conclusions and outlook}
In this paper, we have compared the performance of three different protocols in the presence of state preparation errors and operational errors. The three protocols are: non-catalytic EC, catalytic EC and distillation. To do this,  we have extended the analysis of EC protocols to mixed states and included operational errors. We also propose a new method for obtaining POVMs, allowing for a trade-off between the number of communication rounds and required auxiliary qubits given certain system capabilities.

Through our analysis, we find catalytic EC to be beneficial over distillation if the depolarisation error in the prepared state is small (below 5 percent). We have described how to implement EC protocols on a circuit-based quantum system for obtaining Bell states and analysed the effect of operational errors on the performance of the protocols. We find that catalytic EC requires operational error rates of $\lesssim0.1$\% due to the relatively high complexity of the POVM measurement circuits. This makes catalytic EC relevant for the generation of logical Bell pairs for fault-tolerant, modular quantum computing. In particular, recent work~\cite{pattison2024,ataides2025} has shown that encoding of physical Bell pairs into quantum error correcting codes enables high-rate logical Bell pair distillation. The logical encoding suppresses operational errors as well as the state preparation error of the logical Bell pair well below 0.1\%. Our results show that in this error regime, catalytic EC can achieve both lower infidelity and higher success probability than standard distillation. 

Our current analysis opens up a few interesting future directions. One would be to study the performance of catalytic entanglement concentration for the generation of entangled logical resource states between computing modules in a fault-tolerant modular quantum computer similar to Ref.~\cite{marqversen2025}. A natural extension of our work would be the extension of our methods to the multipartite case. This could be done by extending our methods of finding POVMs to the work in Ref.~\cite{Xin_2007}. However, it remains an open question whether the surrogate-state strategy used here to handle mixed states continues to be effective in the multipartite case. This difficulty is related to the richer structure of multipartite LOCC transformations and entanglement classes~\cite{multipartite_1, Neven_2021, Li2024identifyingfamilies}, which may complicate the identification of suitable reference states.

\begin{acknowledgements}
HS would like to thank Luise Prielinger for her help with the HPC and Nina Codreanu for her help with Adobe Illustrator.
JH acknowledges funding from the Dutch Research Council (NWO) through a Veni grant (grant No.VI.Veni.222.331). JH and HS acknowledge funding from the Quantum Software Consortium (NWO Zwaartekracht Grant No.024.003.037). JB acknowledges support from The AWS Quantum Discovery Fund at the Harvard Quantum Initiative.
Part of this work was performed while HS was on a research visit at Harvard University funded by a Quantum Delta NL travel grant.
\end{acknowledgements}

\section*{Code availability}
The codes for our simulations are publicly available at \url{https://doi.org/10.5281/zenodo.15322618}. The usage guidelines can be found in the Github repository \url{https://github.com/arr0w-hs/catalytic_concentration/}.
\bibliographystyle{quantum}
\bibliography{biblio}
\onecolumn\newpage
\appendix

\section{Local operations and classical communication protocols}\label{sec: locc_intro}
In the context of entanglement distribution, it is generally considered that parties sharing an entangled state can only manipulate their systems and communicate with each other classically. The class of deterministic transformations that satisfy these conditions is called local operations and classical communication (LOCC).

An LOCC transformation is possible if and only if the ordered Schmidt coefficients of the initial and final states follow the majorization condition~\cite{Nielsen_1999}. Consider the initial state $\ket{\psi} = \sum_{i=1}^d \sqrt{\alpha_i} \ket{ii}$ and final state $\ket{\phi} = \sum_{i=1}^d \sqrt{\beta_i} \ket{ii}$, with Schmidt coefficients $\{\alpha_i\}_{i=1}^d$ and $\{\beta_i\}_{i=1}^d$. The Schmidt coefficients of the two states are taken in descending order ($\alpha_i \geq \alpha_j$ if $i<j$). The majorization condition can then be written as:
\begin{align}
    \sum_{i=1}^j \alpha_i \leq \sum_{i=1}^j \beta_i,
\end{align}
for $j<d$ and, $\sum_{i=1}^d \alpha_i = \sum_{i=1}^d \beta_i$. 

LOCC operations impose a constraint on the possible manipulations on shared states~\cite{limitation_of_locc}. However, sometimes if a transformation cannot be done deterministically, it can succeed with a finite probability. Such transformations are called \textit{stochastic} LOCC (SLOCC). Vidal found the protocol for given initial and final states that is optimal in the success probability~\cite{Vidal_1999}. This protocol can be referred to as \textit{conclusive protocol}~\cite{Vidal_2000} because the transformation succeeds with a finite probability but with unit fidelity of the final state,
\begin{align}
    \ket{\psi} \xrightarrow[\text{prob}=p]{\text{fid}=1} \ket{\phi},
\end{align}
where, \textit{fid} is the fidelity of final state, and \textit{prob} is the probability of transformation. 

Vidal's protocol has two phases. The first phase is a deterministic LOCC conversion from the given initial state $\ket{\psi}$ to an intermediate state $\ket{
\gamma}$. The second phase is the application of POVM on the intermediate state $\ket{\gamma}$ to get to the final state $\ket{\phi}$.
\begin{align}
    \ket{\psi} \xrightarrow[\text{prob}=1]{\text{fid}=1} \ket{\gamma}\xrightarrow[\text{prob}=p]{\text{fid}=1} \ket{\phi}.
\end{align}
In his paper, Vidal also introduced a way to find the required operations for the  state transformation. These operations depend on both the the initial state and the final state. Therefore, in general, application of these operations on the incorrect initial state might lead to an incorrect final state. This makes extending these protocols from pure states to mixed states difficult.

\section{Finding the operations for LOCC}\label{sec: our_method}
In this appendix, we discuss our method for finding POVMs such that the number of required auxiliary qubits and rounds of communication can be tuned based on the system's capabilities. Our method is a combination of the method proposed by Nielsen~\cite{Nielsen_1999} (referred to as \textit{Nielsen's method}), and the method proposed by Jensen and Schack~\cite{simple_locc} (we call this the \textit{JS-method}). Our current analysis is limited to bipartite states, however, our protocol can easily be extended to multipartite states following the extension in Ref.~\cite{Xin_2007}. 

We will first briefly introduce Nielsen's and the JS-method for finding the POVM set. We will then discuss how we combine the two previously known methods.

The majorization condition proposed by Nielsen for LOCC can equivalently be written as follows: the entanglement transformation $\ket{\psi} \rightarrow \ket{\phi}$ is possible if and only if a doubly-stochastic matrix $D$ exists such that $V_\psi = D V_\phi$. Where $V_\psi^T = (\alpha_1, ..., \alpha_d)$ and $V_\phi^T = (\beta_1, ..., \beta_d)$ are the vectors constructed from the Schmidt coefficients of the initial and final states, respectively. $D$ is known as a doubly-stochastic matrix. It is a $d \cross d$ square matrix composed of non-negative real numbers such that all of its rows and columns sum to 1.

Nielsen constructed the operations by decomposing the $d \cross d$ doubly-stochastic matrix $D$ into at most d-1 \textit{T-transforms}~\cite{Nielsen_1999}. Where a T-transform is the following: it acts on a vector by preserving all but two elements of the vector. Those two elements are acted upon by the matrix.

\begin{align*}
\begin{bmatrix}
t & 1-t \\
1-t & t \\
\end{bmatrix},
\end{align*}
where, $t \in [0,1]$. This decomposition can be found using the algorithm given in~\cite{rajendra_matrix}. An observation about T-transforms is that it is a convex combination of an Identity matrix and a permutation matrix. Moreover, a T-transform is also a doubly-stochastic matrix, and multiplication of two T-transforms of the same dimensions is also a doubly stochastic matrix. 

Corresponding to each T-transform, a POVM can be constructed with two elements and correction unitaries corresponding to each element can be found. For the execution of LOCC, one of the parties measures the POVM and communicates the result to the other parties. All the parties then apply a correction unitary based on the measurement outcome. This is repeated sequentially for each T-transform. Therefore, the number of communication rounds required by this method is equal to the number of T-transforms in the decomposition of $D$.

Alternatively, the JS-method uses the Birkhoff-von Neumann theorem (chap 2 of~\cite{rajendra_matrix}) to decompose the doubly-stochastic $D$. According to this theorem, a $d\cross d$ doubly-stochastic matrix can be decomposed into a convex combination of at most $(d-1)^2+1$ permutation matrices~\cite{rajendra_matrix}. 

The JS-method creates a set of POVM with elements corresponding to each permutation matrix. Moreover, each permutation matrix is also used to find the correction unitary. The LOCC protocol is similar to Nielsen's method: one of the parties measures the POVM on their system and communicates the result to other parties. Following this, all the parties apply the correction unitary based on the measurement outcome. Due to the construction of the POVM, only a single round of communication is required for this method. However, the POVM set can have at most $(d-1)^2+1$ elements and correction unitaries.

The number of required auxiliary qubits for POVM measurements is one for Nielsen's method and at most $\left\lceil \log_2(d^2 -2d+2) \right\rceil$ for the JS-method. On the contrary, the JS-method requires one round of communication, and Nielsen's method requires d-1 communication rounds. Here, $d$ is the dimension of the doubly-stochastic matrix. 

We combine the two aforementioned methods to find the operations for entanglement concentration. We first find the $d \cross d$ doubly-stochastic matrix ($D$) that relates the vectors of the ordered Schmidt coefficients. We then find the decomposition of  $D$ into T-transforms $D = T_{k}\cdots T_1$, such that $k\leq d-1$. Using this decomposition, we can group and multiply the T-transforms of the decomposition to create doubly-stochastic matrices $D^{new} = T_aT_b \cdots T_c$ such that,
\begin{align}
    D = D_l^{new} \cdots D_1^{new},
\end{align}
 $ l < k$. Here, we have used the property that the multiplication of T-transforms of the same dimension results in a doubly-stochastic matrix. This reduces the number of rounds of communication from k to l. Furthermore, we then apply Birkhoff's theorem to these newly created doubly-stochastic matrices ($D_i^{new}$) to obtain permutation matrices corresponding to each doubly stochastic matrix, using the JS-method. Moreover, it can be deduced that the number of elements in the decomposition of $D_i^{new}$ into permutation matrices requires a lower number of permutation matrices than the decomposition of $D$. This is because a T-transform is a convex combination of an Identity and a permutation matrix. Therefore, $D_i^{new}$s can be created such that we can reduce the number of rounds of communication compared to Nielsen's method while requiring fewer auxiliary qubits than the JS-method.

\section{Execution of POVM}\label{sec: povm_impl}

For the execution of LOCC and SLOCC protocols, it is considered that parties can implement a Positive operator-valued measurement (POVM). A POVM is a set of positive semi-definite Hermitian matrices $\{A_i\}_{i=1}^m$ such that $\sum_{i=1}^m A_i = I$, where $I$ is the Identity matrix. To study the effect of operational errors on our protocols, we convert POVMs into projective measurements on a high-dimensional Hilbert space using Naimark's dilation~\cite{naimark, naimark_peres}. This conversion can be implemented by introducing additional auxiliary qubits to the system. The POVM elements can then be "embedded" into a unitary that operates on the data qubits and the auxiliary qubits. Finally, the auxiliary qubits undergo projective measurements implementing a POVM on the data qubits. This conversion requires $\left\lceil \log_2 m \right\rceil$ number of auxiliary qubits, where $m$ is the number of elements in the set of POVMs.

The core of our LOCC implementation uses the algorithm of the JS-method for obtaining POVMs for multiple rounds of communication. At the beginning of each round, the initial state and final state (for that round) are used to find the POVMs. These POVMs are diagonal matrices with non-negative entries by construction. We exploit this construction for embedding the obtained POVMs into a unitary matrix. However, due to limitations imposed by software simulators, we only consider one auxiliary qubit for POVM measurement. Therefore, we always have POVM with two elements in them that are diagonal matrices with positive entries. An element of our POVM has the following form:
\begin{align}
    A_i = \sum_{j=0}^{k-1} a_i^j \ketbra{j},
\end{align}

such that $i \in \{0,1\}$, and, $a_i^j$ are non-negative real numbers such that $a_0^j + a_1^j = 1$ for each $j \in \{0, ..., k-1\}$. This follows from the completeness condition of POVMs. The value of $k$ depends on the initial state at the start of each round. However, it is always upper-bounded by the dimension of the system $l$:  $k \leq l$. Where the dimension of the system depends on the number of data qubits $n$ (including the catalyst state) as $l = 2^n$. We can therefore construct unitaries $U^j$s from entries in the POVM elements such that, 
\begin{align}
 U^j = \begin{cases} 
      \sqrt{a_0^j} Z +\sqrt{a_1^j} X ,& 0 \leq j\leq k-1 \\
      I ,& k-1 < j< l \\
   \end{cases}
\end{align}
where $j,k,l \in \mathbb{Z} $.

We can then embed these unitaries into a higher-dimensional unitary called the \textit{embedding unitary} that acts on both data qubits and auxiliary qubits. 
\begin{align}
    U_{da} = \sum_{j=0}^{l-1} \ketbra{j} \otimes U^j.
\end{align}

For the case when $k=l$, $U_{da}$ acts on the state $\ket{\psi}_{da} = \ket{\psi}_d\ket{0}_a $ such that:
\begin{align}
    U_{da}\ket{\psi}_{da} = & U_{da} \left(\ket{\psi}_d\ket{0}_a \right)\\
    =& \sum_{j=0}^{l-1} \bra{j}_d \psi\rangle_d \ket{j}_d   U^j \ket{0}_{a}\\
    =& \sum_{j=0}^{l-1} \bra{j}_d \psi\rangle_d \ket{j}_d   (\sqrt{a_0^j} \ket{0}_{a} + \sqrt{a_1^j}\ket{1}_a)
\end{align}

Here, the subscripts $d$, $a$, and $da$ represent data qubits, auxiliary qubits and the combined data-auxiliary system, respectively. Projective measurement of the auxiliary qubit in computational basis, using measurement operators $M^i = \ketbra{i}$ ($i \in \{0,1\}$), results in one of the following states:
\begin{align}
    \ket{\psi}_{da}^m = 
    \begin{cases}
        \sum_{j=0}^{l-1}  \bra{j}_d \psi\rangle_d \sqrt{a_0^j} \ket{j}_d   \ket{0}_{a}  ,& m = +1\\
        \sum_{j=0}^{l-1}  \bra{j}_d \psi\rangle_d \sqrt{a_1^j} \ket{j}_d   \ket{1}_{a}  ,& m = -1\\
    \end{cases}
\end{align}
It can be checked easily that this is equivalent to the application of $A_0$ and $A_1$ on the data qubits, conditioned on the measurement result. Moreover, the probability of success for each measurement would be:
\begin{align}
    p_i =& \bra{\psi_{da}} I_d\otimes M^i_a \ket{\psi_{da}}\\
    =& \sum_{j,k} \bra{j} \psi\rangle_d \bra{\psi} k\rangle_d  \sqrt{a_i^ja_i^k}  \bra{k}
    {j}\rangle_d\\
    =& \sum_{j}  |\bra{j} \psi\rangle_d|^2 (a_i^j)  \\
    =& \sum_{j}  \bra{\psi}_{d} a_i^j \ketbra{j} \psi\rangle_{d}\\
    =&  \bra{\psi}_{d} A_i  \ket{\psi}_{d},
\end{align}
where, $M^i_a$ is the projective measurement operator on the auxiliary qubits. As shown here, the probability of projective measurements on the auxiliary qubits is equivalent to the POVM measurement of the data qubits. 

Our method here can be extended to the case when POVMs have more elements than 2. In such a case, $\left\lceil \log_2 m \right\rceil$ auxiliary qubits would be required such that each projective measurement on them can be mapped equivalently to a POVM element, where $m$ is the number of elements in the POVM. Moreover, the unitary $U^j$ would be a unitary that would act on all the auxiliary qubits. Therefore, $U_{da}$ would be composed of multi-controlled multi-target (MCMT) unitaries with controls being the data qubits and targets being the auxiliary qubits.

\section{POVM synthesis and operational errors}\label{sec: povm_synthesis}

We have two (three) Bell pairs in our simulations of non-catalytic EC (catalytic EC). Each Bell pair requires one data qubit at (for example) Alice's and Bob's labs. Moreover, they both need one more auxiliary qubit to execute the POVM measurements using the embedding unitary. To estimate the effect of operational errors on catalytic EC and non-catalytic EC protocols, we decompose the embedding unitary into single and multi-qubit gates. 

The embedding unitary given by,
\begin{align}
    U_{da} &= \sum_{j=0}^{l-1} \ketbra{j} \otimes U^j, \\
\end{align}
and it can be decomposed into a product of multi-controlled unitaries $U_j'$,
\begin{align}
    U_j' = \ketbra{j} \otimes U^j + \sum\limits_{\substack{i=0 \\ i \neq j}}^{l-1} \ketbra{i} \otimes I.
\end{align}

The unitary $U_j'$ acts on the auxiliary qubits with the unitary $U^j$ if the data qubits are in the state $\ket{j}$, otherwise, it acts with just the Identity. Such a multi-controlled unitary can be synthesised in terms of 3 single qubit gates and 2 multi-controlled X (MCX) gates~\cite{Barenco_1995, Shende_2006}. MCX gates are native to quantum computing platforms like Neutral atoms~\cite{cczgate, Dlaska_2022}. For estimating the effect of operational errors on the fidelity of catalytic EC (non-catalytic EC), we assume that each gate depolarises the qubits it acts on. Moreover, we assume that single-qubit gates are error-free. We make this assumption because multi-qubit gates have an order-of-magnitude higher infidelities compared to single-qubit gates~\cite{cczgate}, and the number of single and multi-qubit gates in our implementation is nearly the same. 

\begin{figure*}[t]
    \centering
    \begin{minipage}{0.85\textwidth}
        \includegraphics[width = 1\linewidth]{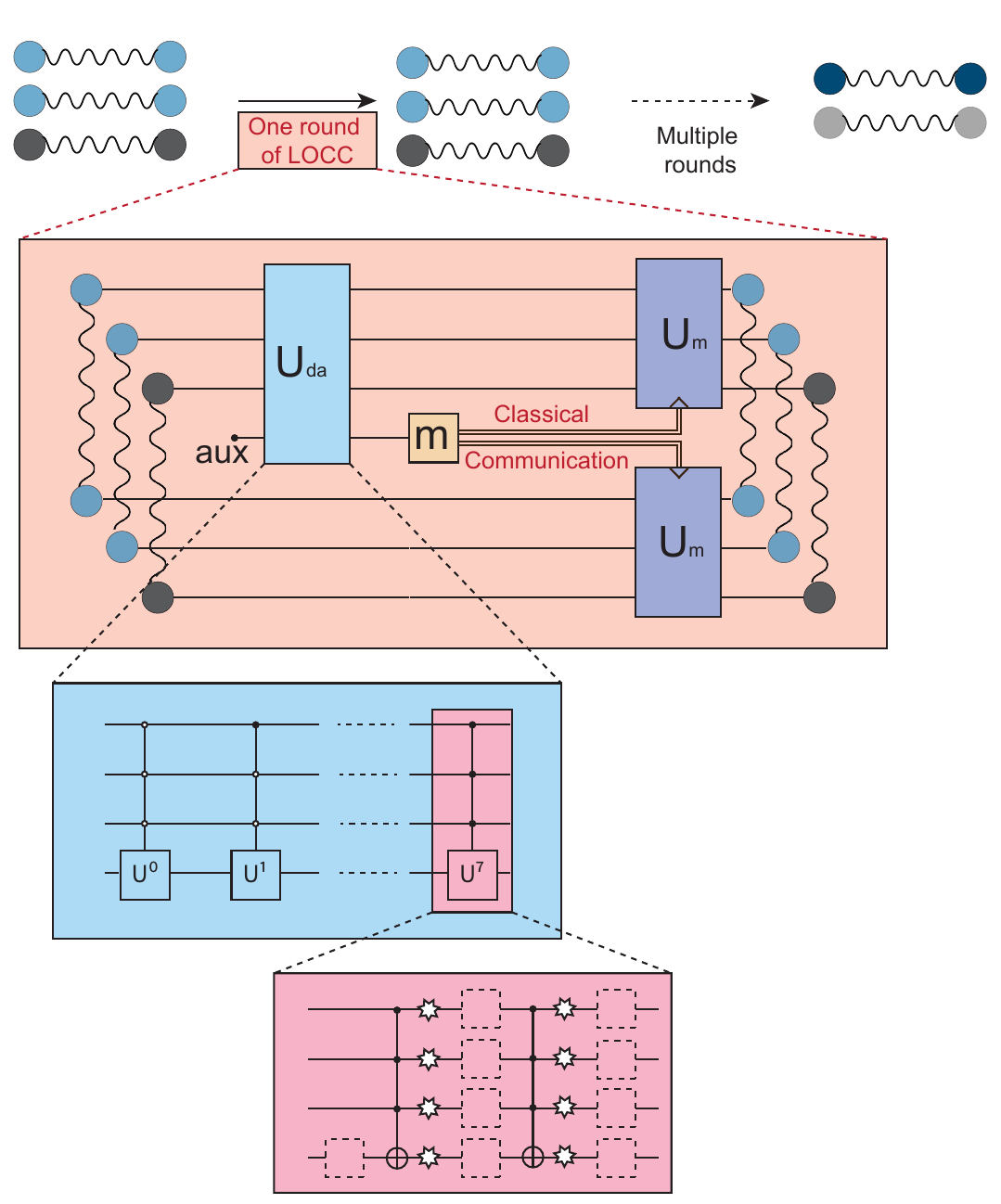}
    \end{minipage}
    \caption{Scheme for executing catalytic EC. It involves the conversion of two weakly entangled states into one highly entangled state using a catalyst state. This involves multiple rounds of local operations and classical communications. A round of LOCC involves the application of embedding unitary $U_{da}$ followed by measurement of the auxiliary qubit in the computational basis. This is then followed by the application of correction unitaries $U$ depending on the measurement result. We synthesise $U_{da}$ into a series of multi-controlled unitaries $U'_j$ that act on the auxiliary qubit with $U^j$ if the data qubits are in the $\ketbra{j}$ state. Each unitary $U'_j$ can be synthesised using at most two MCX gates, as shown in~\cite{Shende_2006}. For modelling operational errors, we assume that each MCX gate depolarises each qubit it acts on, shown using the white stars. The scheme for non-catalytic EC is the same except we do not have the catalyst state there, thus, the number of qubits involved is fewer than catalytic EC.}
    \label{fig: state_prep2}
\end{figure*}

In our implementation of catalytic EC and non-catalytic EC, each round of POVM measurement involves the execution of $U_{da}$ unitary for that round. This unitary can further be written as a product of unitaries $U'_j$. These unitaries can be represented in terms of at most two MCX gates, shown in Fig.~\ref{fig: state_prep2}. We assume that whenever an MCX gate acts on a set of qubits, it depolarises the qubits slightly by a given value of error rate. Thus, each communication round involves the application of multiple MCX gates. We note that this way of synthesising $U'_j$s into multiple MCX gates might not be optimal because of the exact values of the unitaries. But it is sufficient for us to find an upper bound on the operational errors.

Other than the POVM unitary, each communication round involves the application of correction unitaries that correspond to the measurement result of the POVMs. In the JS-method, these correction unitaries are permutation matrices that permute the bases. Thus, we can keep track of these matrices and implement them classically for changing the required bases. Since these can be implemented classically, we can omit the application of these unitaries, and thus, they do not cause any operational errors.

\end{document}